\documentclass[nocite]{epl}
\newcommand{\be}{\begin{equation}}
\newcommand{\ee}{\end{equation}}
\newcommand{\ba}{\begin{eqnarray}}
\newcommand{\ea}{\end{eqnarray}}
\usepackage{color}

\usepackage{graphicx}
\usepackage{dcolumn}
\usepackage{bm}
\usepackage{epsfig}
\usepackage{subfigure}
\begin{document}



\title{ On the rigidity of a hard sphere glass near random close packing}


\author{Carolina Brito\inst{1,3}, Matthieu Wyart\inst{2,3}}
\institute{\inst{1} Instituto de F\'{\i}sica, Universidade Federal do
 Rio Grande do Sul, 91501-970 Porto Alegre, Brazil \\
\inst{2} Division of Engineering and Applied Sciences, 
Harvard University, Pierce Hall,
29 Oxford Street, Cambridge, Massachusetts 02138, USA\\
\inst{3}CEA - Service de Physique de l'Etat Condens\'e,
Centre d'Etudes de Saclay, 91191 Gif-sur-Yvette, France }

\date{\today}
\maketitle

\begin{abstract}
We study theoretically and numerically the microscopic cause of the rigidity of hard sphere glasses near their maximum packing. We show that, after coarse-graining over time, the hard sphere interaction can be described by an effective potential which is exactly logarithmic at the random close packing $\phi_c$. This allows to define normal modes, and to apply recent results valid for elastic networks:  rigidity is a non-local property of the packing geometry, and is characterized by some length scale $l^*$ which diverges at $\phi_c$ \cite{matthieu1,matthieu2}.  We compute the scaling of the bulk and shear moduli near $\phi_c$, and speculate on the possible implications of these results for the glass transition.
\end{abstract}

Hard spheres present a glass phase between $\phi_0$,  where the glass transition
occurs and structural relaxation becomes unobservable, and $\phi_c$ where the
pressure $p$ diverges. In this region this system is solid and resists to shear on any measurable time scales. Although a large amount of works focused on the super-cooled liquid, the glass itself received less attention.  In particular, there is no undoubted microscopic theory to explain its mechanical properties and its rigidity.  In the cage-escape picture \cite{sjorgen}, the cage formed by the neighboring particles tighten as $\phi$ increases,  and the typical time for a particle to escape its cage grows and eventually diverges. Nevertheless,  Maxwell showed that the stability against {\it collective} motions of particles is more demanding than against individual particle displacements: in particular $z=d+1$ inter-particle contacts are sufficient to pin one particle in $d$ dimensions, whereas $z_c=2d$ contacts {\it in average} are required to guarantee mechanical stability \cite{max}. Thus considering {\it a priori} rigidity as a local property  may be inappropriate. 

Recently several works \cite{matthieu1,matthieu2,ohern,J,these,head}
studied the mechanical properties of {\it weakly-connected} elastic
networks  with an average contacts number ---the coordination number--- $z$ close to the critical
value $z_c=2d$, such as those encountered for athermal repulsive
short-range particles {\it above} $\phi_c$ \cite{ohern,J}. In
particular it was shown that (i) these systems present an excess of vibrational modes at low-frequency in comparison with normal solids \cite{J}. These {\it anomalous modes} are characterized by some length $l^*\sim\delta z^{-1}$
\cite{matthieu1,matthieu2}, where 
$\delta z\equiv z-z_c$,  (ii) rigidity can occur only if  $\delta z\geq C_0 \sqrt{p/B}$  
on any subsystems of size $l\geq l^*$, where $C_0$ is a constant  and $B$ is the  bulk modulus \cite{matthieu2}.  Thus
rigidity is a {\it non-local} property of the packing geometry,  (iii)  the shear modulus $G$ satisfies $G/B\sim \delta z$, 
as observed numerically \cite{J} and confirmed theoretically  \cite{these} for repulsive systems. 

Can these results apply to hard sphere glasses? 
On the one hand,  hard spheres are weakly-connected at high packing fraction, being exactly $z=z_c$ at $\phi_c$ as was shown theoretically \cite{shlomon,moukarzel,tom1,roux,donev,zamponi}.
On the other hand,   all the results established for elastic networks require
 a smooth potential to expand the energy and define normal modes. It
 is in principle  problematic in hard sphere systems where the potential is
 discontinuous.  In this Letter we show that, once a coarse-graining
 in time is made, hard spheres interact with a continuous
 effective potential, which becomes exactly logarithmic as  $\phi\to
 \phi_c$. This allows  to define normal modes
 and to derive new results on the rigidity and the mechanical responses of the glass near $\phi_c$.

\begin{figure}[htbp]
\begin{center}
\rotatebox{0}{\resizebox{4.5cm}{!}{\includegraphics{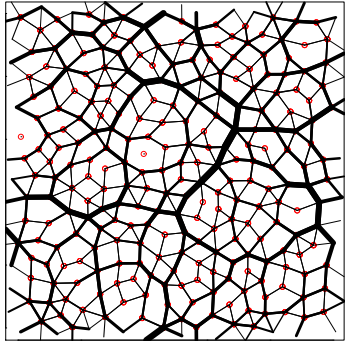}}}
\caption{Contact forces for $N=256$,
  $\epsilon=10^{-4}$ and $t_1 = 10^5$ collisions. Particles centers are represented by red points and the contact forces  by the segments whose width  is proportional
  to the force amplitude. 
 Note that the forces are balanced within our data precision on every particle, as it
 must be the case on  time scales where the structure is stable.
For similar force networks see \cite{donev2}. }
\label{forcefield}
\end{center}
\vspace{-0.5cm}
\end{figure}

We consider a hard sphere glass, where particles collide elastically,  at  high packing fractions $\phi$ close to $\phi_c$ where structural relaxation is frozen. 
The particle diameter defines the unit length. Since temperature only rescales the time unit we fix $\beta=1$.
Following \cite{bubble,donev,donev2} it is possible to define a {\it contact force network}. We introduce an
 arbitrary time $t_1$ much larger than the collision time $\tau_c$. Two particles are said to be {\it in contact} if they
 collide with each other during a time interval  of length $t_1$. This allows to define a coordination number $z\equiv2N_c/N$, where
 $N_c$ is the total number of contacts and $N$ is the particles number. Then, the contact
 force $\vec{f}_{ij}$ between two particles $i$ and $j$ is defined as the total momentum
they exchange  per unit time:
\be
\label{imp}
\vec{f}_{ij}=\frac{1}{t_1}\sum_{n=1}^{n=n_{col}[t_1]} \Delta \vec{P}_n ,  
\ee
where the sum is made on the total number of collisions $n_{col}[t_1]$
between $i$ and $j$ that took place in the time interval $t_1$, and $\Delta \vec{P}_n$ is the momentum exchanged
at the nth shock. Fig(\ref{forcefield})  shows a two-dimensional  example of the
 contact force network obtained with a polydisperse configuration
\footnote{Half of the particles
 have a diameter unity, the other half as a diameter 1.4.} 
at packing fraction $\phi$  close to $\phi_c$. 
To obtain high packing fractions  numerically we used the 2-dimensional jammed
 configurations of \cite{J} with packing fraction  $\phi_c\approx 0.83$. At  $\phi_c$ the particles
 are in contact. We reduce the particles diameters
 by a relative amount $\epsilon$.  This leads to configuration of
 packing fraction $\phi=\phi_c(1- \epsilon)^2$.  Then, we assign a random 
 velocity to every particle and launch an event-driven   simulation. 
Such system is not at thermal equilibrium and displays aging \cite{brito}:  ``earthquakes'' can occur  which suddenly relax the system and decrease the pressure \footnote{ Similar earthquakes have been observed in other aging systems such as colloidal pastes, laponite or Lennard-Jones simulations \cite{aging}. They correspond to a sudden collective rearrangement of a large number of particles.}. In between these rare events, there are very long quiet periods where no structural relaxation is observed.
All our measures are done during these periods. Note that coordination and contact forces could a priori 
depend on the arbitrary parameter $t_1$. In the vicinity of $\phi_c$ no significant dependence
of these quantities with $t_1$ were observed as long as (i) $\tau_c\ll t_1$ and (ii) no earthquake occurs in the time interval $t_1$.

As we shall see below, the force networks of dense hard sphere glasses are weakly-connected.  
To build a correspondence between hard spheres at $\phi<\phi_c$ 
and elastic  spheres at $\phi>\phi_c$, we change
variables: instead of considering the instantaneous particles positions
we consider their time-average position
$\{\vec{R}^{av}_i\}$  
over some time scale $t_1\gg\tau_c$. 
To define an effective potential we must relate the contact force $f_{ij}$ to the average
distance  between particles $i$ and $j$. We start by considering 
the simple example of a line of hard spheres equilibrated at some pressure $p$. The isothermal isobaric partition function is easily computed by introducing the {\it spacings} $h_i$  between particles, defined as
 $h_i=r_{i,i+1}-r_i-r_{i+1}$, where $r_{i,i+1}$ is the distance
 between particles $i$ and $i+1$, and $r_i$ is the radius of particle $i$. One obtains:
 \be
 {\cal Z} \sim \prod_i\int_{h_i=0}^{h_i=\infty} dh_i e^{- \beta p h_i}, 
 \ee
where the terms containing the kinetic energy of the particles have been integrated out. If an external 
force dipole $p_i=-p_{i+1}\equiv p_1$ is now applied on $i$ and $i+1$, the work required to open the contact $i$ of an amount $h_i$ is now $(p+p_1) h_i$. Thus we obtain:
  \be
 {\cal Z}\sim \prod_{j\neq i}\int_{h_j=0}^{h_j=\infty} dh_j e^{-\beta p h_j}   \int_{h_i=0}^{h_i=\infty} dh_i e^{-\beta (p+p_1) h_i}.
 \ee
 It is then straightforward to compute the average spacing $\langle h_i\rangle=1/\beta(p+p_1)$.  Since the contact force $f_i$ in the contact $i,i+1$ is $f_i=p_1+p$, one finds that all contacts satisfy the relation:
 \be
 \label{ff}
 f_j=\frac{1}{\beta\langle h_j\rangle} \ \ \ \ \ \ \hbox{ for all } \ j
 \ee
which is thus true whether external forces are present or not.

\begin{figure}[htbp]
\begin{center}
 \rotatebox{-90}{\resizebox{7.0cm}{!}{\includegraphics{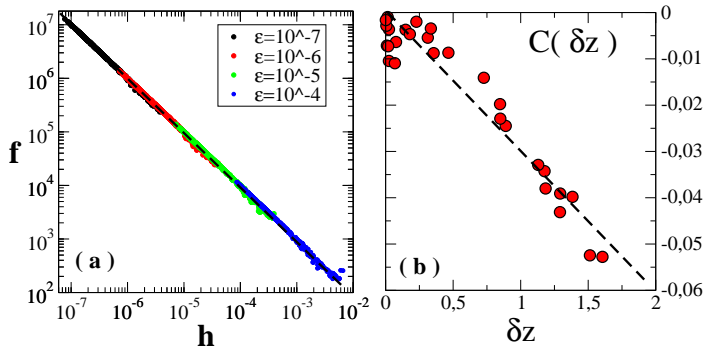}}}
  \vspace{-2.0cm}
    \caption{{\small (a) --  Log-log plot of the  contact force amplitude
         {\it vs.}  the spacing  $h=r-r_i-r_j$ for various $\epsilon$ in  systems of  $N=256$
         particles in two dimensions. Each dot represents the  pair of numbers
         ($f_{ij}$,  $\langle h_{ij}\rangle$) associated with the contact
         $ij$.  Dots collapse  on the dotted theoretical curve defined by
         Eq.(\ref{ff}).   (b) -- Average correction $ C(\delta z)$  as defined
         in the text    {\it vs.} excess coordination $\delta   z$ for
         various $\phi$.   The line is a linear fit  consistent with the
         predictions of  \cite{these} at small $\delta z$. Corrections are
         small, of the order of 3 to 4 percent when $\delta z=1$.}}
\label{force_and_correction} 
\end{center}
\vspace{-0.5cm}
\end{figure}

We now demonstrate that Eq.(\ref{ff}) can be extended to hard sphere glasses
at $\phi=\phi_c$ for any spatial dimension. As discussed above, at $\phi_c$,
if the ``rattlers''   are removed
\footnote {At $\phi_c$ some particles ($\approx 5 \%$)   do not have any 
contact, and lie at a finite distance of their neighbors. These 
``rattlers''  do not participate to the rigidity of the structure. 
Near $\phi_c$ these particles can be  identified  since
the distance with their neighbors is much larger than the average. In 
all our measures we defined ``rattlers" as the particles for which
their second   strongest contact force is smaller than 1\% of 
the system-averaged contact force. 
We checked that our results, such as the scaling of 
the coordination and the pressure, are still valid when other definitions of
rattlers are used (for  example taking a threshold of 5\% instead of 1\% 
for the contact 
force).}
, the system is marginally connected, or {\it isostatic}: $z=z_c$.  Isostatic states have the particularity to display as many degrees of freedom of displacements as number of contacts. Hence, (i) the configuration of the system can be defined  by the set of distances between particles in contact and (ii) these degrees of freedom are independent. This implies that the isobaric partition function is a product of terms corresponding each to an individual contact. Consequently, if the system is at equilibrium in a meta-stable state where the  contact forces field $|{\bf f}\rangle = \{ f_{ij}\}$ is well-defined, the isobaric partition function can be written \footnote{The upper limits of the integrals of Eq.(\ref{part}) are not infinite, but bounded by some finite value $h_{max}$ which depends of the contact considered. Nevertheless,
 as in the one-dimensional case, $h_{max}\sim N/\langle f \beta\rangle $
 \cite{donev}. Since the integrals in Eq.(\ref{part}) converge as soon as
 $h\gg 1/\beta f$,  Eq.(\ref{part}) becomes exact when $N>>1$ for any $\phi$ near $\phi_c$.}:
 \be
 \label{part}
 {\cal Z}\sim\prod_{ \langle ij \rangle} \int_{h_{ij}=0}^{h_{ij}=\infty} dh_{ij} e^{-\beta f_{ij} h_{ij}}.
 \ee
Repeating the argument valid for $d=1$, one obtains that $f_{ij}=\langle
 h_{ij}\rangle^{-1}\beta^{-1}$. Obviously, as is the case in one dimension,
 this result is valid with or without external forces.   Note that since this
 derivation only invokes thermodynamic arguments, it also applies to Brownian
 particles. This relation  force/distance  is checked numerically in
 Fig.(\ref{force_and_correction}-a)  near $\phi_c$.  The  dependence of $f$ with $h$ is found to be in very good agreement with Eq.(\ref{ff}). 

When $\phi$ is lowered from $\phi_c$, we shall see that the coordination $z$ increases. Then, the $h_{ij}$ are not independent
 variables anymore in Eq.(\ref{part}) and Eq.(\ref{ff}) is invalid. Nevertheless the relative corrections to Eq.(\ref{ff}) are expected to be small, of order $\delta z$ \cite{these}. 
 We check this result in Fig.(\ref{force_and_correction}-b), where we compute numerically $C(\delta z)\equiv
 \langle f_{ij}\beta \langle h_{ij}\rangle\rangle_{ij}-1$, where
 $\langle\rangle_{ij}$ denotes the average over all contacts.   In what follows,  we are mainly interested in scaling relations near $\phi_c$, 
 for which corrections of  order $\delta z$ are not relevant. We shall negle
ct them.
 
In this approximation, it is straightforward to compute the thermodynamic potential from Eq.(\ref{part}) (or by integrating Eq.(\ref{ff})): ${\cal G}= - \beta^{-1} \sum_{\langle ij\rangle} \ln(\langle h_{ij}\rangle)$. A key remark is that this expression of ${\cal G}$ corresponds precisely to the  energy of an athermal assembly of particles of positions  $\{\vec{R_i}^{av}\}$, interacting with a smooth potential $V_{ij}$ of the form:
\ba
\label{pot}
V_{ij}(r)&=&\infty \hspace{3.5cm}  \hbox{  if} \ \ \ \ r<r_i+r_j  \nonumber \\
V_{ij}(r) &=& - \beta^{-1}\ln (r-r_i-r_j)  \hspace{0.8cm}  \hbox{ if $i$ and $j$ are in ``contact''} \nonumber  \\
V_{ij}(r) &=& 0  \hspace{3.7cm}  \hbox{ if $i$ and $j$ are not in ``contact"}
\ea
where $r\approx || \vec{R}^{av}_i-\vec{R}^{av}_j||$ is the average distance
between $i$ and $j$. This analogy between hard spheres and systems with soft
interactions implies that: (i) a configuration minimum of the  
thermodynamic potential is also a minimum of the energy in the corresponding
elastic system, and must therefore satisfy the rigidity criterion evoked in
introduction.(ii) when a hard sphere system is sheared (or compressed), the
change of thermodynamic potential  can be deduced from the
shear (bulk) modulus of the elastic system. (iii) Since the effective
potential of Eq.(\ref{pot}) is  continuous the thermodynamic potential can be expanded around any configuration.  This allows to compute the dynamical matrix $\cal M$ \cite{Ashcroft} and the normal modes defined as the eigenvectors of $\cal M$. 

We now precise (i) to derive a microscopic criterion for the rigidity, or meta-stability, of dense hard sphere glasses. 
Any meta-stable state must contain at least one configuration corresponding to a minimum of thermodynamic potential. 
Using (i), this implies that for this configuration $\delta z> C_0 \sqrt{p/B}$ \cite{matthieu2}. Anticipating on
 what follows,  we have for hard spheres $B\sim p^2$ and $p\sim (\phi_c-\phi)^{-1}$, therefore there is a constant $C_1$ such that:
\be
\label{dz}
\delta z \geq C_1 p^{-1/2}\sim (\phi_c-\phi)^{1/2}.
\ee 
This is our main result, which relates rigidity and microscopic structure. 

\begin{figure}[htbp]
\begin{center}
\vspace{-0.0cm}
\rotatebox{-90}{\resizebox{8.3cm}{!}{\includegraphics{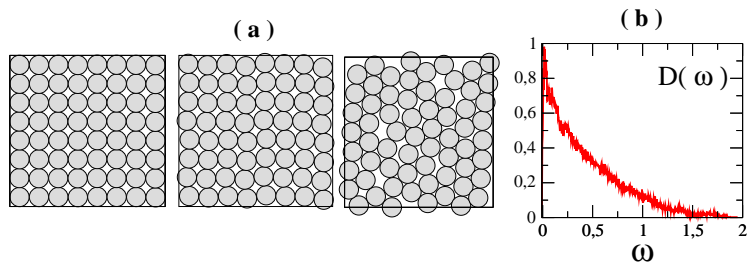}}}
\vspace{-4.0cm}
\caption{{\small (a) Collapse of a square lattice.
(b) $D(\omega)$ {\it vs.} $\omega$ for $\epsilon=10^{-4}$ in
 a poly-disperse glass.
 All frequencies are rescaled by $\epsilon^{-1}$. The particle positions
  were averaged over a time  $t_1 = 4 * 10^5 $ to obtain $|{\bf  R}^{av}\rangle$.  Eq.(\ref{pot})  was used to compute the dynamical matrix, from which the normal modes frequencies were inferred.}}
\label{Dw3}
\end{center}
\vspace{-0.5cm}
\end{figure}

To test this prediction we study three different systems:  
the two-dimensional hard sphere glass introduced 
above,  the mono-disperse crystal and the mono-disperse square lattice.  We consider all these systems at their maximum packing fraction where particles are in permanent
 contact, then we reduce the particles diameter by a relative amount
 $\epsilon$, and we launch a simulation. 
 In the crystal case, the coordination is 6, therefore $\delta
 z=2\gg p^{-1/2}\sim \epsilon^{1/2}$ for small $\epsilon$: condition (\ref{dz}) is satisfied  and  the system is stable. 
 On the other hand,  the square configuration has  $z=4$, $\delta z=0$,  and the system cannot satisfy
 (\ref{dz})  without large structural rearrangements for any  $\epsilon$.
These predictions are verified numerically. For small $\epsilon$, the crystal
 is stable and displays not structural changes, whereas the square lattice collapses
 rapidly \cite{shlomon, donev2}, see Fig.(\ref{Dw3}(a)). 

\begin{figure}[htbp]
\begin{center}
\rotatebox{-90}{\resizebox{4.5cm}{!}{\includegraphics{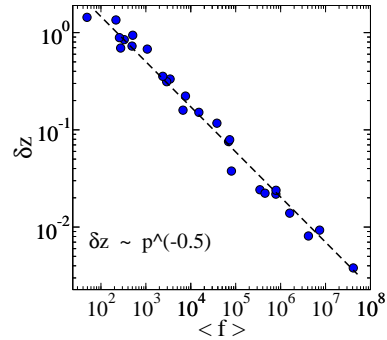}}}
\caption{{\small Log-log plot of $\delta z$  versus the average 
contact force $\langle f\rangle\sim p$. The data were obtained for
 different $\epsilon$ and different time periods.
The black line corresponds to the equality of the inequality (\ref{dz}). } }
\label{PvsZ}
\end{center}
\end{figure}

As discussed above, the poly-disperse glass we obtain near $\phi_c$ presents long periods of stability, where no relaxation occurs.
To  check Eq.(\ref{dz}) we computed numerically both the coordination and the pressure for  various packing
fractions, and for various stable periods that appear along the aging
regime.  As shown in Fig.(\ref{PvsZ}), the data are consistent with an {\it equality} of the inequality (\ref{dz}).
This suggests that  a hard sphere glass lies close to {\it marginal} stability, as is the case for soft spheres slowly decompressed toward $\phi_c$ \cite{matthieu2}.

In Fig.(\ref{Dw3}-b) we also furnish an example of density of states
$D(\omega)$  for $\epsilon=10^{-4}$ computed during such a plateau, when the
``rattlers'' are removed.
Note that $D(\omega)$ does not vanish as $\omega\to 0$, as in any isostatic
system \cite{matthieu1}.  It is the case here,  
since $\phi$ is very close to $\phi_c$, and therefore $\delta z\approx 0$. Interestingly no unstable modes
are observed at this packing fraction. This is not obvious: in a
meta-stable state, the system could lie alternatively in several minima of the
thermodynamic potential, since the temperature is non-zero. Condition (\ref{dz}) would then be satisfied, as is the case for each minimum.  Nevertheless, after averaging, the position may lie in between several minima, near a saddle-point where unstable modes are present. This situation may well occur at lower $\phi$. We leave this question and its possible relation with some observed structural relaxation processes \cite{sjorgen} for future investigations. 

To compute the scaling of the elastic moduli of the glass near $\phi_c$ we can use the results valid
 for repulsive weakly-connected elastic networks. For the bulk modulus one obtains \cite{J,these}:
\be
\label{mm}
B\sim \frac{\delta p}{\delta \phi}\sim (\phi_c-\phi)^{-2}.
\ee
as found previously \cite{donev,zamponi}. The same scaling holds for the crystal \cite{frenkel}.
As discussed in introduction, in repulsive weakly-connected network the shear modulus does not scale as $B$, but rather as $G\sim B \delta z$.
 Making the assumption that in the glass phase, the system does not depart
 much from marginal stability, Eq.(\ref{dz}) is an equality, and one obtains the new result:
\be
\label{gg}
G\sim  p^{3/2} \sim (\phi_c-\phi)^{-3/2}.
\ee

To conclude, we showed that an analogy can be made between a hard sphere glass and an elastic system once a coarse graining in time is made, and 
that the effective potential that describe particle interactions becomes exactly logarithmic  at $\phi_c$ where the pressure diverges.
This allows to define normal modes, and to compute the scaling of the elastic moduli near $\phi_c$. 
This implies that the rigidity that characterizes the glass phase near maximum packing is related to a {\it non-local}
microscopic property of the system geometry: the {\it coordination number} $z$ must be bounded below on any subsystems 
larger than some length $l^*\sim \delta z^{-1}$ which diverges at $\phi_c$. 
Finally, our numeric data suggest that the glass phase is only marginally stable, at least in the vicinity of $\phi_c$,
 implying the presence of anomalous modes near zero-frequency.

One may question if these results apply in the vicinity of the glass transition. On the one hand, near $\phi_0$  the coordination is rather large, similar to the one of the crystal.  One the other hand, the distribution of contacts stiffness  is certainly broader in the glass. This enhances the presence of anomalous modes at low-frequency, since softer contacts affect only weakly the vibrational spectrum \cite{these}.  Thus, anomalous modes could be an appropriate concept to study how rigidity appears when $\phi$ increased toward $\phi_0$. In particular, it has been proposed that the dramatic slow down near $\phi_0$  corresponds to a transition in the topology of the free-energy landscape \cite{parisi,laloux}: at high $\phi$, the system lies near free-energy minima, and the dynamics is activated. At lower $\phi$, the system lives near saddle-points, and the dynamic consists in going down the unstable directions of the free-energy. Our work suggests the following hypothesis: these unstable directions correspond to anomalous modes. Since such modes are collective particles motions, they may cause the heterogeneous dynamics  \cite{Ediger} observed near the glass transition.

We thank JP Bouchaud, F. Leonforte, S. Nagel and T. Witten for helpful discussions, and L. Silbert for furnishing the jammed configurations. C.B. was financed by 
CNPq  and  CAPES.

\end{document}